\documentclass[twocolumn]{svjour3}          
\smartqed  
\usepackage[english]{babel}        
\usepackage{graphicx}
\usepackage{amssymb}
\usepackage{amsmath}
\usepackage[usenames,dvipsnames]{color}
      \newcommand{\etal}{et~al. }
      
 \journalname{arXiv.org}			
\begin{document}

\title{Experimental determination and modelling of volume shrinkage in curing
thermosets}



\author{  Martin Rudolph \and 
          Martin Stockmann \and
	  Ralf Landgraf \and
          J\"orn Ihlemann
}	

\institute{M. Rudolph \and
             M. Stockmann \and
             R. Landgraf \and
             J. Ihlemann  \at
              Department of Solid Mechanics, \\
              Institute of Mechanics and Thermodynamics, \\
              Faculty of Mechanical Engineering, \\
              Chemnitz University of Technology, 
              Chemnitz, Germany \\[1mm]
              M. Rudolph \at 
              \email{martin.rudolph@mb.tu-chemnitz.de} 
              \\[1mm]
              M. Stockmann \at \email{martin.stockmann@mb.tu-chemnitz.de}
              \\[1mm]
              R. Landgraf \at \email{ralf.landgraf@mb.tu-chemnitz.de}
              \\[1mm]
              J. Ihlemann \at \email{joern.ihlemann@mb.tu-chemnitz.de}
}

\maketitle

\begin{abstract}
This work deals with the characterisation and modelling of the curing process and its associated volume changes of an epoxy based thermoset resin. Measurements from differential scanning calorimetry (DSC) define the progress of the chemical reaction. The related thermochemical volume changes are recorded by an especially constructed experimental setup based on Archimedes principle. Information on measuring procedure and data processing are provided. This includes investigations on compensation of environmental influences, long-term stability and resolution. With the aim of simulating the adhesives curing process, constitutive models representing the reaction kinetics and thermochemical volume changes are presented and the model parameters are identified.

\end{abstract}


\newbox\JIGAMMa
\newbox\JIGAMMb
\newbox\JIGAMMc
\newbox\JIGAMMd
\newbox\JIGAMMz
\newbox\LLbox
\newbox\LLboxh
\newbox\SLhilfbox
\newbox\SLubox
\newbox\SLobox
\newbox\SLergebnis
\newbox\TENbox
\newif\ifSLoben
\newif\ifSLunten
\newdimen\JIGAMMdimen
\newdimen\JIhsize\relax\JIhsize=\hsize
\newdimen\SLrandausgleich
\newdimen\SLhoehe
\newdimen\SLeffbreite
\newdimen\SLuvorschub
\newdimen\SLmvorschub
\newdimen\SLovorschub
\newdimen\SLsp
\def\rhotilzurho{{\JI\frac{\STAPEL\varrho!^\SLtilde\!}{\JI\!\varrho}}}
\def\rhozurhotil{{\JI\varrho\over\JI\STAPEL\varrho!^\SLtilde}}
\def\pkt{\cdot}
\def\ppkt{\mathbin{\mathord{\cdot}\mathord{\cdot}}}
\setbox\JIGAMMa = \hbox{$\scriptscriptstyle c$} \setbox\JIGAMMz =
\hbox{\hskip-.35pt\vrule width .25pt\hskip-.35pt
                        \vbox to1.2\ht\JIGAMMa{\vskip-.125pt
                             \hrule width1.2\ht\JIGAMMa height.25pt
                             \vfill
                             \hrule width1.2\ht\JIGAMMa height.25pt
                             \vskip-.125pt}%
                        \hskip-.125pt\vrule width .25pt\hskip-.125pt}
\def\Oldroy#1#2#3{\STAPEL{#1}!_\SLstrich!_\SLstrich!^\circ{}
                  \ifx #2,{}_{\copy\JIGAMMz}%
                  \else \mskip1mu{}^{\copy\JIGAMMz}\fi
                  \mskip1mu\ifx #3,{}_{\copy\JIGAMMz}%
                           \else {}^{\copy\JIGAMMz}\fi }
\def\OP#1#2{\ifnum#1=1{\rm S}
            \else\ifnum#1=2{\rm S}^\JIv
                 \else\ifnum#1=3{\rm S}^T
                      \else{{\rm S}^T}^\JIv
            \fi\fi\fi\LL{{#2}}\RR}
\edef\JIminus{{\setbox\JIGAMMa=\hbox{$\scriptstyle x$}%
           \hbox{\hskip .10\wd\JIGAMMa
                 \vbox{\hrule width .6\wd\JIGAMMa height .07\wd\JIGAMMa
                       \vskip.53\ht\JIGAMMa}%
                 \hskip .10\wd\JIGAMMa}}}
\edef\JIv{{\JIminus 1}}
\def\JI{\displaystyle}
\def\JIha{{1\over 2}}
\def\JIfolgt{\quad\Rightarrow\quad}
\def\LL#1\RR{\setbox\LLbox =\hbox{\mathsurround=0pt$\displaystyle
                                              \left(#1\right)$}%
       \setbox\LLboxh=\hbox{\mathsurround=0pt%
                  $\displaystyle{\left(%
                      \vrule width 0pt height\ht\LLbox depth\dp\LLbox
                      \right)}$}%
       \left(\hskip-.3\wd\LLboxh\relax\copy\LLbox
              \hskip-.3\wd\LLboxh\relax\right)}
\def\ZBOX#1#2#3{\def#3{}%
                \setbox#1 = #2
                \def#3{ to \wd#1}%
                \setbox#1 = #2}
\def\SLdreieck{\setbox\TENbox=\hbox{\fontscsy\char 52}
                 \dp\TENbox = 0pt
                 \hbox{\hskip -2\SLrandausgleich
                       \box\TENbox
                       \hskip -2\SLrandausgleich}}
\def\SLtilde{\setbox\TENbox=\hbox{\fontscex\char 101}
                    \vbox{\vskip-.03\ht\TENbox
                          \hbox{\hskip -1\SLrandausgleich
                                \copy\TENbox
                                \hskip -1\SLrandausgleich}
                          \vskip -.86\ht\TENbox}}
\def\SLstrich{\vrule width \SLeffbreite height.4pt}
\def\SLpunkt{{\vbox{\hbox{$\displaystyle.$}\vskip.03cm}}}
\def\SLabstand{\vskip .404pt}
\def\SLzwischen{\vskip 1.372pt}
\font\fontscsy=cmsy6 \font\fontscex=cmex10 scaled 1200
\def\STAPEL#1{\def\SLkern{#1}%
              \futurelet\next\SLpruef
               A_0   _0    :B_0   _-.17 :C_.05 _-.15 :D_0   _-.2
              :E_0   _-.2  :F_0   _-.21 :G_0   _-.15 :H_0   _-.23
              :I_.2  _.15  :J_.05 _-.1  :K_0   _-.22 :L_0   _-.1
              :M_0   _-.23 :N_0   _-.25 :O_.05 _-.2  :P_0   _-.21
              :Q_.05 _-.2  :R_0   _-.03 :S_0   _-.15 :T_.2  _0
              :U_.1  _-.1  :V_.1  _-.15 :W_.1  _-.2  :X_0   _-.22
              :Y_.16 _-.15 :Z_0   _-.25
              :a_.05 _-.05 :b_.05 _0    :c_.05 _.05  :d_0   _-.05
              :e_.07 _0    :f_0   _-.15 :g_.04 _-.2  :h_0   _-.07
              :i_.05 _0    :j_.08 _-.1  :k_0   _-.1  :l_.2  _.15
              :m_0   _-.1  :n_0   _-.1  :o_0   _-.1  :p_.15 _0
              :q_.1  _0    :r_.1  _-.1  :s_0   _-.2  :t_.1  _.05
              :u_0   _-.1  :v_0   _-.2  :w_0   _-.2  :x_.04 _-.14
              :y_.15 _-.05 :z_0   _-.15
              :\mit\Phi_.08 _-.1    :\mit\Omega_0 _-.2   :\varXi_.00 _-.2
              :\alpha_0 _-.2        :\gamma_.1 _-.1      :\varepsilon_.1 _-.1
              :\epsilon_.05 _-.05   :\eta_.05 _-.15      :\lambda_0 _0
              :\mu_0 _-.25          :\nu_0 _-.2          :\varSigma_-.03 _-.2
              :\varrho_.00 _-.2     :\sigma_.1 _-.2      :\tau_.15 _-.15
              :\theta_.2 _0          :\nabla_.1 _-.2
              :\varphi_.2 _-.1      :\omega_.1 _-.1      :\mit\Gamma_-.1 _-.1
              :\Lambda_0 _0         :\Gam_0 _0           :\Lam_0 _0
              :{\cal K}_1 _0
              :\SLsuchende
              \def\SLtrick{\noexpand\SLtrick\noexpand}%
                \def\SLdummy{\noexpand\SLdummy}%
                \edef\SLoboxinhalt{}\edef\SLuboxinhalt{}%
                \SLobenfalse\SLuntenfalse
                \futurelet\next\SLsuchruf}
  \def\SLsuchruf{\ifx\next !\let\next\SLexpand
                 \else\let\next\SLerzeug\fi\next}
  \def\SLexpand#1#2#3{\ifx #2\sb\ifSLunten\let\SLspeicher\SLuboxinhalt
                   \else\def\SLspeicher{\SLtrick\SLabstand}\fi
                   \edef\SLuboxinhalt{%
                       \SLspeicher
                       \SLtrick\SLzwischen
                       \hbox\SLdummy{\hfil\mathsurround=0pt
$\SLtrick\scriptstyle\SLtrick#3$%
                                     \hfil}}%
                   \SLuntentrue%
                      \else\ifSLoben\let\SLspeicher\SLoboxinhalt
                   \else\def\SLspeicher{\SLtrick\SLabstand}\fi
                   \edef\SLoboxinhalt{%
                       \hbox\SLdummy{\hfil\mathsurround=0pt
$\SLtrick\scriptstyle\SLtrick#3$%
                                     \hfil}%
                       \SLtrick\SLzwischen
                       \SLspeicher}%
                   \SLobentrue\fi\futurelet\next\SLsuchruf}
  \def\SLerzeug{\def\SLtrick{}
                \setbox\SLhilfbox=\hbox{$\displaystyle{E}$}%
                \SLrandausgleich=.04\wd\SLhilfbox
                      \setbox\SLhilfbox=%
                         \hbox{\hskip -1\SLrandausgleich
                          \mathsurround=0pt$\displaystyle{\SLkern}$%
                               \hskip -1\SLrandausgleich}%
                      \SLhoehe = \ht\SLhilfbox
                      \advance\SLhoehe by \dp\SLhilfbox
                      \SLeffbreite = \wd\SLhilfbox
                      \advance\SLeffbreite by \SLab\SLhoehe
                      \ZBOX\SLubox{\vbox{\offinterlineskip
                                         \SLuboxinhalt
                                         \hrule height 0pt}}\SLdummy
                      \ZBOX\SLobox{\vbox{\offinterlineskip
                                         \SLoboxinhalt
                                         \hrule height 0pt}}\SLdummy
                      \SLsp = \SLzu\SLhoehe
                      \advance\SLsp by -.5\SLeffbreite
                      \SLuvorschub = -1\SLsp
                      \advance\SLuvorschub by -.5\wd\SLubox
                      \SLovorschub = -1\SLsp
                      \advance\SLovorschub by -.5\wd\SLobox
                      \advance\SLovorschub by .26\SLhoehe
                      \ifdim\SLuvorschub > \SLovorschub
                         \SLsp = \SLovorschub
                      \else
                         \SLsp = \SLuvorschub
                      \fi
                      \ifdim\SLsp < 0pt%
                         \advance\SLuvorschub by -1\SLsp
                         \SLmvorschub = -1\SLsp
                         \advance\SLovorschub by -1\SLsp
                      \else
                         \SLmvorschub = 0pt
                      \fi
                      \setbox\SLergebnis = \hbox{%
                         \offinterlineskip
                         \hskip\SLrandausgleich\relax
                         \vbox to 0pt{%
                            \vskip -1\ht\SLobox
                            \vskip -1\ht\SLhilfbox
\hbox{\hskip\SLovorschub\copy\SLobox\hfil}%
                            \hbox{\hskip\SLmvorschub\copy\SLhilfbox
                                  \hfil}%
\hbox{\hskip\SLuvorschub\copy\SLubox\hfil}%
                            \vss}%
                         \hskip\SLrandausgleich}%
                      \SLsp = \dp\SLhilfbox
                      \advance\SLsp by \ht\SLubox
                      \dp\SLergebnis = \SLsp
                      \SLsp = \ht\SLhilfbox
                      \advance\SLsp by \ht\SLobox
                      \ht\SLergebnis = \SLsp
                      \box\SLergebnis{}}
  \def\SLpruef{\ifx\next\SLsuchende\def\SLzu{0}\def\SLab{0}%
                  \def\next##1\SLsuchende{\relax}%
               \else\let\next\SLvergl
               \fi\next}
  \def\SLvergl#1_#2_#3:{\def\SLv{#1}%
                        \ifx\SLkern\SLv\def\SLzu{#2}\def\SLab{#3}%
\def\next##1\SLsuchende{\relax}%
                        \else\def\next{\futurelet\next\SLpruef}
                        \fi\next}
\def\PKT#1{#1!^\SLpunkt}

\newbox\minusbox
\def\minus{\mathchoice{\minusarb\displaystyle}%
                      {\minusarb\textstyle}%
                      {\minusarb\scriptstyle}%
                      {\minusarb\scriptscriptstyle}}
  \def\minusarb#1{\setbox\minusbox=\hbox{$#1x$}%
                  \hbox{\hskip .10\wd\minusbox
                        \vbox{\hrule width .6\wd\minusbox
                                     height .07\wd\minusbox
                              \vskip.53\ht\minusbox}%
                        \hskip .10\wd\minusbox}}

\def\Basis{\STAPEL e!_\SLstrich}
\def\C{\STAPEL C!_\SLstrich!_\SLstrich}
\def\X{\STAPEL X!_\SLstrich!_\SLstrich}
\def\Cinv{\STAPEL C!_\SLstrich!_\SLstrich^{-1}}
\def\Cg{\STAPEL C!_\SLstrich!_\SLstrich!^\SLstrich}
\def\CD{\STAPEL C!_\SLstrich!_\SLstrich!^\SLdreieck}
\def\XD{\STAPEL X!_\SLstrich!_\SLstrich!^\SLdreieck}
\def\ECKMT{\STAPEL M!_\SLstrich!_\SLstrich!^\SLdreieck^T}
\def\CgD{\STAPEL C!_\SLstrich!_\SLstrich!^\SLstrich!^\SLdreieck}
\def\BD{{\STAPEL C!_\SLstrich!_\SLstrich!^\SLdreieck}{^{\JIv}}}
\def\BDn#1{{\STAPEL C!_\SLstrich!_\SLstrich!^\SLdreieck}{^{\JIv}_{#1}}}
\def\Ttil{\STAPEL T!_\SLstrich!_\SLstrich!^\SLtilde}
\def\rhotil{\JI\STAPEL\varrho!^\SLtilde\!}
\def\rhotilzurho{{\JI\frac{\STAPEL\varrho!^\SLtilde\!}{\JI\!\varrho}}}
\def\rhozurhotil{{\JI\varrho\over\JI\STAPEL\varrho!^\SLtilde}}
\def\Skal#1{\STAPEL {#1}}
\def\Vek#1{\STAPEL {#1}!_\SLstrich}
\def\Ten2#1{\STAPEL {#1}!_\SLstrich!_\SLstrich}
\def\F{\STAPEL F!_\SLstrich!_\SLstrich}
\def\Fg{\STAPEL F!_\SLstrich!_\SLstrich!^\SLstrich}
\def\e{\STAPEL e!_\SLstrich!_\SLstrich}
\def\b{\STAPEL b!_\SLstrich!_\SLstrich}
\def\D{\STAPEL D!_\SLstrich!_\SLstrich}
\def\I{\STAPEL I!_\SLstrich!_\SLstrich}
\def\I{\STAPEL I!_\SLstrich!_\SLstrich}
\def\L{\STAPEL L!_\SLstrich!_\SLstrich}
\def\P{\STAPEL P!_\SLstrich!_\SLstrich}
\def\CnOldWed{\overset{\circ{}}{\Big(\STAPEL{C}!_\SLstrich!_\SLstrich_2\Big)}
                  \mskip1mu\mskip1mu
                  \mskip1mu {}^{\hat{\copy\JIGAMMz}}
                  \mskip1mu {}_{\hat{\copy\JIGAMMz}}}
\def\CnOldWedred{\STAPEL{C}!_\SLstrich!_\SLstrich!^\circ{}_2}

\def\CD{\C!^\SLdreieck}
\def\CDn#1{\C!^\SLdreieck_#1}
\def\X{\Ten2 X}
\def\XDn#1{\X!^\SLdreieck_#1}
\def\Cauchy{\STAPEL \sigma!_\SLstrich!_\SLstrich}
\def\Cn#1{\C_{#1}}
\def\Ln#1{\L_{#1}}
\def\h{\STAPEL h!_\SLstrich!_\SLstrich}
\def\Fn#1{\F_{#1}}
\def\qtil{\STAPEL q!_\SLstrich!^\SLtilde}
\def\nabtil{\STAPEL \nabla!^\SLtilde!_{\hspace{0.5ex}\SLstrich}}
\def\omtil{\STAPEL \omega!_\SLstrich!^\SLtilde}
\def\curetil{\STAPEL q!^\SLtilde}
\def\thetatil{\STAPEL \theta!^\SLtilde}
\def\Ktil{\STAPEL {\cal K}!^\SLtilde}
\def\Vtil{\STAPEL V!^\SLtilde}
\newcommand{\IntGtil}[1]
   {\displaystyle{{\Big.^{\widetilde{\mathcal{G}}}}
      \hspace{-1.5ex}\int #1 \ {\rm d}\widetilde{V}}}
\newcommand{\IntGtilN}[2]
   {\displaystyle{{\Big.^{\widetilde{\mathcal{G}}_{#1}}}
      \hspace{-1.5ex}\int #2 \ {\rm d}\widetilde{V}_{#1}}}
\newcommand{\IntGhat}[1]
   {\displaystyle{{\Big.^{\widehat{\mathcal{G}}}}
      \hspace{-1.5ex}\int #1 \ {\rm d}\widehat{V}}}
\newcommand{\nadel}[1]
   {\big(\hspace{-3pt}\big(\hspace{1pt} #1\hspace{1pt}\big) \hspace{-3pt}\big)}
  
\newcommand{\inv}{{\minus 1}}
\newcommand{\invT}{{\minus\T}}    

\def\indLT#1#2%
  {\mathop{}%
   \mathopen{\vphantom{#2}}^{\scriptscriptstyle #1}%
   \kern-\scriptspace%
   #2}
%
\section{Introduction}
\label{sec:Introduction}

Against the backdrop of the increasing demand for light\-weight construction, the importance of adhesives to a wide range of applications in all fields of engineering increases steadily. Especially in the case of joining different material classes, many conventionel methods, like welding (for metals) or friction welding (for plastics) fail to apply.
Since methods like screwing, rivetting and clinching cause high local stresses to a component and in addition can represent a starting point to cracks, often the joining by adhesives is given the preference.
Within the last years, the world wide production of adhesives has significantly increased. 
 The fields of application has historically been enlarged and spans nowadays not only aviation, automotive and railway engineering but also the fields of mechanical engineering, eletronics and mechatronics \cite{Habenicht2006}. 
A very famous application of adhesives is the construction of fibre-reinforced compound materials, such as e.g. ARALL and GLARE.
%
%
It can easily be seen that there is a wide range of applications but, however, all these fields have in common that they are subjected to one main disadvantage of this material class: i.e. the accompanying volume shrinkage during the curing reaction. This probably might be the most challenging aspect to practical processing of adhesives.
There are chemically and technologically ways to minimise this effect and, hence, reduce the concomitants like risk of component distortion, formation of cracks, delamination and residual stresses. Nevertheless, the stated effects do not vanish by any of these efforts.

Consequently, there arises a need for its reliable measurement and prediction to cope with curing related shrinkage.
Toward this end, \cite{DaSilva2011} 
 gives an comprehensive overview on applicable methods. Most of them are limited to linear dimension changes or only account for the total volume shrinkage at the end of the hardening. However, if e.g. the production process of lightweight components is to be simulated, like in \cite{Landgraf2014}, \cite{Mahnken_2013} 
 or \cite{Drossel_Etal_2009_CIRP}, 
 exact knowledge of the adhesives properties throughout the process is vital. When it comes to real-time measurement, the fact that there is a phase change to the material from liquid to solid within the curing process limits the number of possible methods for detection. Here, methods based on the immersion principle are most promising. In~\cite{DaSilva2011} 
 both the temperature control as well as the handling of very small volume changes are pointed out as crucial aspects to conduct a well defined experiment.

Different works have dealt with the monitoring of volume changing effects based on Archimedes principle, see for example \cite{Li2004}, \cite{Kolmeder_Lion_2010}, \cite{Khoun2010}, and \cite{Lion_Yagimli_2008}. 
A detailed literature review on this topic is given in~\cite{Nawab2013}. 

Within this article, thermochemical volume changes of an epoxy resin are considered. Both experimental characterisation as well as simulative results based on phenomenological modelling are presented.
First of all, the curing process itself is characterised and modelled in section~\ref{sec:Curing_reaction}. As pointed out in \cite{Halley_Mackay_1996} 
measurements from differential scanning calorimetry (DSC) are suitable to define the hardening progress. Here, the measuring principle and a modelling approach for the simulation of curing kinetics are outlined.
Next, an especially constructed experimental setup based on Archimedes principle is presented in section~\ref{sec:Modelling_Curing_Shrinkage}. It enables for the real-time measurement of volume and density changes of arbitrary materials. The setup includes two essential features, i.e. the compensation of environmental influences via a dummy signal and the combination of two different measuring systems.
 More precisely, changes in buoyancy force are transformed into deflections of a spring beam and into equivalent displacements of a connected rigid bar. The applied combination of two measurement systems (strain gauges and capacitive sensors) with different measuring ranges allows for simultaneous detection of both the absolute volume as well as thermochemically related, comparably small changes of this volume.
Investigations on resolution, long-term stability and the compensation method are given subsequently.
Moreover, a procedural protocol to the real-time measurement as well as mathematical relations for the calculation of the density from measured signals are deduced.
Finally, the basic measure for the representation of the curing kinetics is employed to model the thermochemically related changes in density or volume. A comparison between experimental data and simulation data is given in the end.

%
\section{Curing reaction} 
\label{sec:Curing_reaction}

It is vital for comparison of measurement and modelling of the chemically related volume shrinkage to characterise the curing reaction itself in first place. Since polymers have been widely studied within the last decades, they are well understood nowadays. This is both concerning the applied experimental methods for characterisation as well as for the modelling approaches. Thus, only a brief summary of the underlying aspects with corresponding literature references is presented below.
The polymer which is studied within in this article is an epoxy resin based two component adhesive. It is a cold-hardening system and hence a thermoset with a comparably low glas transition temperature. The ongoing reaction is a polyaddition. In~\cite{DP410_2003} indications for the duration of the curing reaction are included.

\subsection{Reaction kinetics}
\label{sec:DSC}

In order to monitor the reaction kinetics of an epoxy, several methods can be applied. Dynamic scanning ca\-lo\-ri\-me\-try (DSC), fourier transform infrared spec\-tros\-co\-py (FTIR), dielectric measurements and rheokinetic measurements are pointed out in \cite{Halley_Mackay_1996} 
 and discussed in detail. In \cite{Hulder2008} 
 the DSC as an heat flow method is ascertained to be the most widely employed technique to characterise the progress of the curing reaction. Hence, the curing process has been measured by employing this experimental method analogously to the approaches described in \cite{Kolmeder_Lion_2010} and \cite{Lion_Yagimli_2008}. The reaction enthalpy can be monitored by the evaluation of the heat flow difference between a reference and a sample containing pan (with a mass of about~$5-8\,$mg) over time while following a prescribed temperature profile. First, a linear temperature ramp is applied to the specimen in a so called dynamic DSC-experiment. From this experiment the ultimate reaction enthalpy~$H_{max}$ is obtained, which represents the amount of energy that is necessary to completely convert all of the reactants into the products. Second, isothermal tests at different temperature levels are conducted. Again the heat flow over time is recorded and the enthalpy~$H_{\theta_i}(t)$ for the set temperature~$\theta_i (t)$ at the current reaction time~$t$ is obtained via numerical integration. The ratio of the currently already converted energy amount from the isothermal DSC to the total reaction enthalpy from the dynamic DSC is called degree of conversion, chemical coordinate or degree of cure. It reads as follows:
\begin{equation}
\label{eq:q_from_exp}
\ \ q = \dfrac{H_{\theta_i}(t)}{H_{max}} \quad \text{with} \quad 
        H_{\theta_i}(t) = \int\limits_{0}^{t} \left( \dfrac{\text{d}Q_{\theta_i}}{\text{d}\bar{t}}                                              \right) \text{d}\bar{t} \ ,
\end{equation}

wherein~$\frac{\text{d}}{\text{d}\bar{t}}Q_{\theta_i}$ is the rate of the released reaction heat. Though not explicitly noted, the degree of cure is a scalar time and temperature dependent variable taken values between zero and one for a not and a completely cured material, respectively.
For further details on the evaluation of the recorded measures see e.g. \cite{Landgraf2014} 
 and \cite{Hulder2008}. 
The work presented in this article includes results from one dynamical and five isothermal DSC measurements. 
A comparative illustration between experiment and model is given in section~\ref{sec:DoC}.
%
%
%
%
%
Next, some aspects on the analytical description of reaction kinetics are stated. 

\subsection{Degree of Cure} 
\label{sec:DoC}

\begin{figure}[ht]
	\centering
  \includegraphics[width=0.45\textwidth]{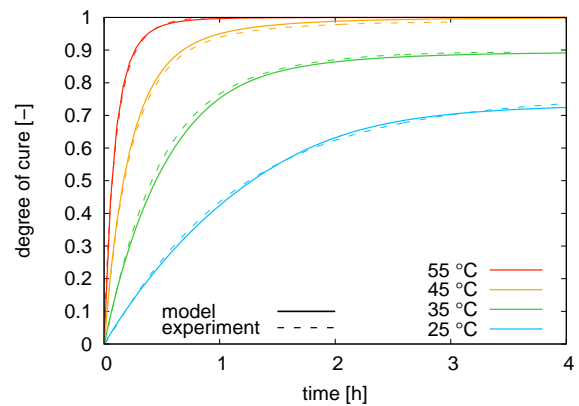}
	\caption{Comparison of DSC measurements and the corresponding modelling approach: evolution of the degree of cure $q$ for different temperatures $\theta$}
	\label{fig:curing}
\end{figure}

As previously mentioned the research on polymers is already very comprehensively explored. It is e.g. in \cite{Nawab2013} 
 and \cite{Halley_Mackay_1996}, 
 where extensive surveys on modelling approach\-es to (inter alia) the reaction kinetics of polymers can be found. Moreover, \cite{Holst2001} 
 provides very detailed information on the phenomenological modelling of the rate equation of the investigated epoxy resin system. The time integration of this equation corresponds the degree of cure. A similar approach to describe the reaction speed of the considered material taken from~\cite{Halley_Mackay_1996} 
 is applied within this article. It is an ordinary differential equation and often referred to as model of the reaction-order type (or nth-order reaction model). 
It reads as follows:
\begin{equation}
\label{eq:q_ansatz}
  \ \ \dfrac{\text{d}q}{\text{d}t}  = K_1(\theta) \cdot\left(1-q\right)^n \cdot f_D(q,\theta) \ ,
\end{equation}
in which~$K_1$ is a proportionality factor with exponential dependence on the curing temperature after Svante A. Arrhenius (cf. \cite{Halley_Mackay_1996}, \cite{Holst2001}) 
\begin{equation}
\label{eq:q_Kfunc}
 \ \ K_1(\theta) = K_{10} \, {\rm exp}\left[-\frac{E_1}{R\,\theta}\right] \ ,
\end{equation}
containing an pre-exponential activation factor~$K_{10}$, the activation energy~$E_1$ and the universal gas constant~$R$. In Eq.~\eqref{eq:q_ansatz} the function~$f_D(q,\theta)$ is also referred to as diffusion factor, since it accounts for the abrupt decrease in reaction speed when the developing glass transition temperature of the material passes the current reaction temperature and thus causes a change to a diffusion-controlled reaction regime. Fournier \etal \cite{Fournier_Etal_1996} proposed an empirical approach 
\begin{equation}
\label{eq:q_diffusion}
 \ \ f_D(q,\theta) = \dfrac{2}{1+{\rm exp}\left[\frac{q-q_{end}(\theta)}{b}\right]}-1 \ ,
\end{equation}
containing a maximum attainable degree of cure~$q_{end}(\theta)$ which depends on the current reaction temperature. It can be calculated by employing the DiBenedetto relation (cf.~\cite{DiBenedetto_1987}) which yields in the accordingly dissolved form:
\begin{equation}
\label{eq:q_qend}
 \ \ q_{end}(\theta) = \dfrac{f(\theta)}{f(\theta)-\lambda \, f(\theta) + \lambda} \ .
\end{equation}
It includes the process dependent glass transition temperature via:
\begin{equation}
\label{eq:q_fTheta}
 \ \ f(\theta) = \dfrac{T_g(q) - T_{g,0}}{T_{g,1}-T_{g,0}} \ 
\end{equation}
into the model. Subsequently, for isothermal processes the assumption $T_g(q) = \theta + \Delta T$ can be employed (cf. \cite{Wenzel2005}): 
Therein, $\Delta T$ denotes the difference between the maximum glass transition temperature $T_g(q_{end}(\theta))$ attainable at specific isothermal curing temperatures, and the curing temperature~$\theta$ itself. The glass transition temperatures~$T_{g,0}$ (uncured material), $T_{g,1}$ (completely cured material) and the parameters~$\lambda$ and~$\Delta T$ can be determined directly from DSC-measurements as described in~\cite{Hulder2008}. 
 The remaining parameters of the model \eqref{eq:q_ansatz} - \eqref{eq:q_diffusion} are identified by application of non-linear optimization methods in conjunction with the DSC measurement results. The identified parameter values are listed in Table~\ref{tab:MatPar_Cure}. A comparison of measurement results and model predictions for the degree of cure at four different temperatures is depicted in Fig.~\ref{fig:curing}.


\begin{table}[ht]
 \centering
  \caption{Material parameters for Eqs.~\eqref{eq:q_ansatz} - \eqref{eq:q_fTheta}} 
  {\begin{tabular}{p{1.6cm}p{2.0cm}p{1.6cm}p{1.4cm}}
    \hline & & & \\[-3mm]
    parameter & value & parameter & value\\ 
    \hline & & & \\[-3mm]
        $\ \ $ $K_{10}$    & $1.608\cdot10^{10} \rm$  
     & $\ \ $ $T_{g,1}$   & $324.85 \ \rm K$   \\
        $\ \ $ $E_1$        & $79835 \ \rm J/mol$         
     & $\ \ $ $T_{g,0}$   & $234.35 \ \rm K$   \\  
        $\ \ $ $b$            & $0.057$                         
     & $\ \ $ $\Delta T$  & $11 \ \rm K$   \\
        $\ \ $ $n$            & $1.217$                          
     & $\ \ $ $\lambda$  & $1.7$   \\ 
    \hline
  \end{tabular}}
  \label{tab:MatPar_Cure}
\end{table}  

%
\section{Thermochemical volume changes}
\label{sec:Modelling_Curing_Shrinkage}

The real-time measurement of chemical shrinkage and heat expansion during the curing process of polymers is a challenging task.
As discussed in \cite{DaSilva2011} 
 the knowledge and control of the temperature is vital to successful investigations. This is both due to the exothermic reaction as well as a result of environmental influences. The former implies the reduction in specimen volume so the generated heat can effectively escape. Since a progress including the materials phase change from liquid to solid is considered, it is not possible to create a geometrically defined specimen. Hence, the density measurement based upon the immersion principle is the only method to be applied. 
If the buoyancy force of the specimen is to be utilized it naturally requires a fluid of lower density. Considering the polymer samples density is already comparably low (cf. \cite{DaSilva2011}: 
 low-density material) 
 there is little scope for a fluid generating a large buoyancy force. Moreover, the expected relative volume shrinkage constitutes only approximately five per cent. Thus, in order to appropriately resolve the expected small force change over time the sample volume should be chosen as large as possible, which is in contrary to the requirements of an isothermal process. Furthermore, the experimental setup should allow for free volumetric changes without the build up of residual stresses.
%
%
To satisfy the specified requirements, an experimental setup based on Archimedes principle has been developed. 
The specific test rig is presented in section~\ref{sec:Principle_measurement_and_exp_setup} and basic investigations on resolution and long-term stability are given in section~\ref{sec:Preliminary_experiments}. In order to identify the volumetric changes due to the curing process at different temperatures, a specific measuring protocol was developed, which is explained in section~\ref{sec:Processing_Shrinkage}. Finally, in section~\ref{sec:model_volume_changes} the test setup is applied for the characterisation of thermochemical volume changes of the considered epoxy resin and a simple phenomenological model is presented to simulate the observed behaviour.

\subsection{Experimental setup} 
\label{sec:Principle_measurement_and_exp_setup}

\begin{figure*}[ht]
  \begin{minipage}[c]{\textwidth}
  \centering
  \includegraphics[width=0.95\textwidth]{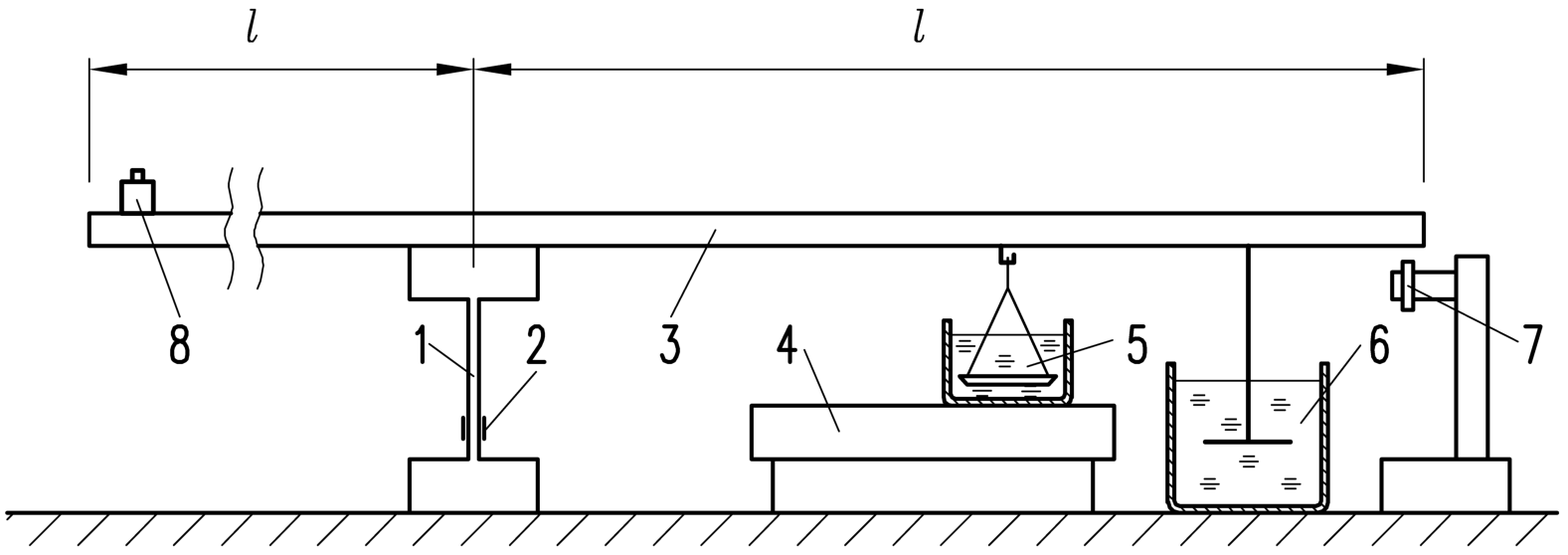}
	\caption{Schematic of the experimental setup for real-time volume shrinkage measurement during the curing reaction consisting of: $1\dots$bending springs, $2\dots$strain gauges, $3\dots$rigid pointers, $4\dots$precision heating plate, $5\dots$specimen on sampler in silicon oil, $6\dots$dashpots, $7\dots$capacitive displacement sensors, $8\dots$balance weights. Except for the specimen~($5$) there are two identical systems in parallel to record a measuring and a dummy signal (see also Fig.~\ref{fig:shrinkage_exp_setup_DMS} and~\ref{fig:shrinkage_exp_setup_cap}). By calculating the difference of these signals environmental effects like temperature changes or air flow are compensated.}
	\label{fig:shrinkage_exp_setup_schematic}
  \end{minipage}
\\[5mm]
  \begin{minipage}[c]{0.47\textwidth}
  \centering
  \includegraphics[width=\textwidth]{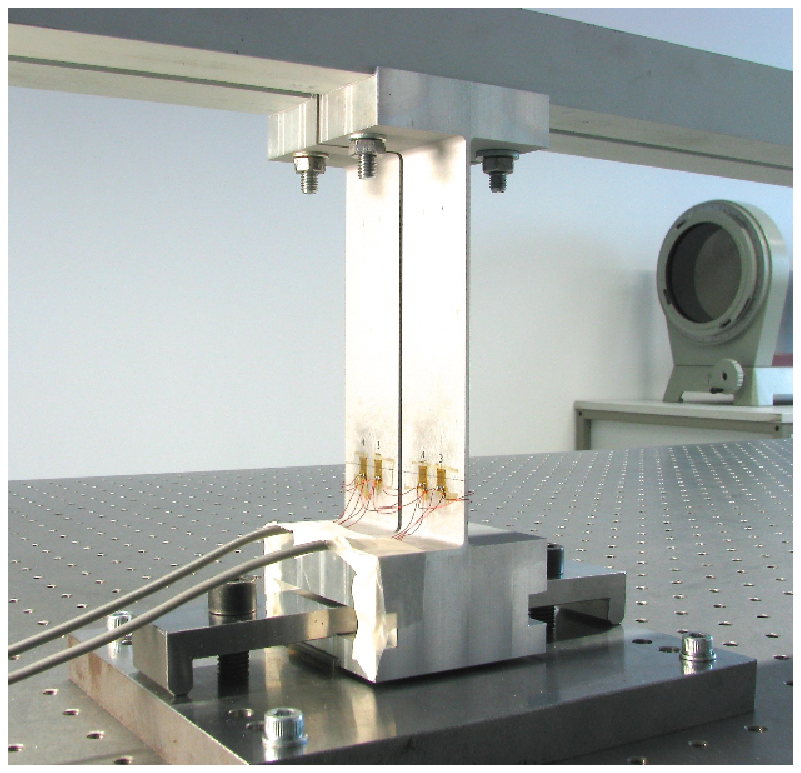}
	\caption{Mounting plate and spring bellow: no interaction between the measuring and a dummy system could be detected. {\color{white}--- dies ist eine dummy-Zeile, da sonst die beiden Abbildungen nicht gleich ausgerichtet sind}}
	\label{fig:shrinkage_exp_setup_DMS}
  \end{minipage}
\hfill
  \begin{minipage}[c]{0.47\textwidth}
  \vspace{0pt}\centering
  \includegraphics[width=\textwidth]{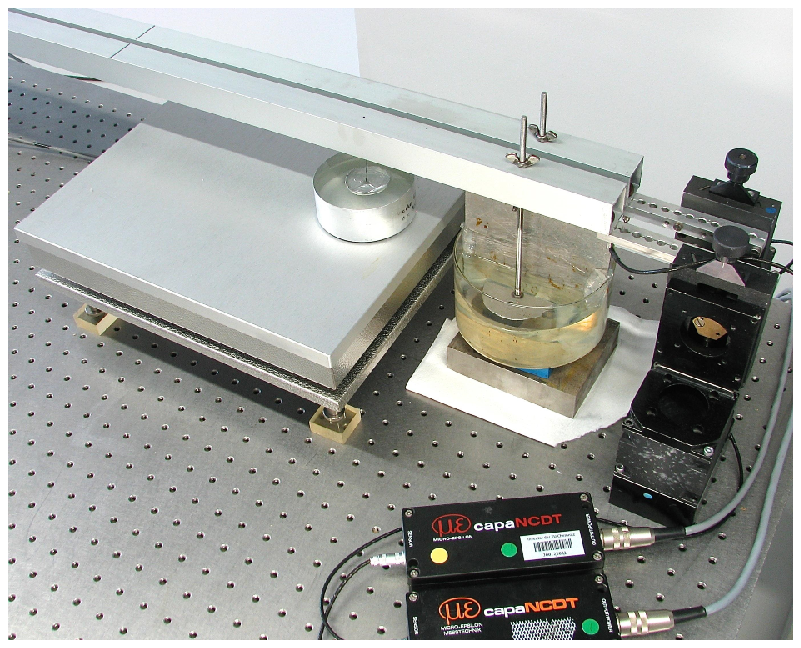}
	\caption{Sampler, precision heating plate, two chamber damping container (dashpots) and adjustable capacitive sensors (isolation and shielding is removed for better visualisation)}
	\label{fig:shrinkage_exp_setup_cap}
  \end{minipage}
\end{figure*}

Figs.~\ref{fig:shrinkage_exp_setup_schematic}-\ref{fig:shrinkage_exp_setup_cap} introduce the experimental setup based on Archimedes principle. In the following the numbers in brackets also refer to those in Fig.~\ref{fig:shrinkage_exp_setup_schematic}.
%
%
%
%
Immersion of the specimen in the testing liquid with known density leads to a
detectable force difference between sample weight and buoyancy, which is subjected to variation as the volume of the specimen changes due to thermochemical processes. In order to monitor this force difference, the setup is equipped with two different measuring systems. One of them is an interconnection of strain gauges to a wheatstone full-bridge
circuit ($2$), which is attached to a vertical bending beam~($1$) with $1.5\,$mm of
thickness. On top of the bending beam a centred rigid aluminium profile~($3$) of
about two metres of length is connected. The movement of this pointer due to changes
in buoyancy force and the resulting varying deflection of the bending beam is
recorded by the second measuring system, a capacitive sensor~($7$) at
the outer end of the profile. Based on the idea presented in~\cite{Nebel2009}, 
 these two different measuring systems are combined in order to detect both absolute density values as well as small changes due to the curing process. This will be considered more detailed in section~\ref{sec:Processing_Shrinkage}.

From the capacitive sensor toward the centre, the hook for the
sampler and a precision heating plate~($4$) carrying the container filled with fluid~($5$)
for the samples immersion is arranged. During measurements, almost the entire space between the two
opposing surfaces of heating plate and measuring system (except the area where
the sampler is hooked to the pointer) is shielded and isolated to keep the
fluids temperature level approximately constant and reduce 
interferences due to heat convection or radiation.

To account for environmental
influences to the measuring system, like changes in room temperature or air
convection and flow, the setup described above exists once again in parallel. Thus, there are two identical systems to record a measuring and a dummy signal. For further evaluation, the difference of these two is formed and thereby the mentioned influences are compensated. No interaction between the two sides of the system could be detected. This principle of compensation works very well for long-term processes as will be demonstrated
in section~\ref{sec:Preliminary_experiments}. 

As a result of high
rigidity and very small masses, the system needs damping. For this purpose, both pointers include a distanced semicircular plate moving
through a fluid (dashpots~($6$)) for viscous damping. It is essential to adjust the fluids
viscosity within a certain range, to not falsify the long-term movements due to
the curing process on the one hand, and to achieve a remarkable damping to
oscillations on the other hand. This aspect is again considered in more detail
in section~\ref{sec:Preliminary_experiments}.

Temperature has decisive influence on the curing behaviour,
effects signals of the strain gauges and changes the density of the fluid for
sample immersion. To this end, thermocouples are
used to detect temperature of the precision heating plate, the room temperature close to the strain gauges at the vertical bending beams and the fluid around the
specimen. Latter was chosen to be a silicon oil due to certain
characteristics like density, viscosity and interaction with the sample. Since the density of both components
of the resin is slightly larger than~$1\,$g$\cdot$cm$^{-3}$ and the curing reaction leads to
further increase in density, a less dense silicon oil was chosen.
The thermal dependency of the fluids density is mandatory to know for evaluating
the samples density like explained in section~\ref{sec:Processing_Shrinkage} and
given by a linear relation (manufacturer information):
\begin{equation}
\label{eq:fluid_density}
   \rho_{fl}(\theta) = \rho_{fl_{ref}} \cdot \left( 1 + \alpha^{th}_{fl} \cdot
(\theta-\theta_{ref}) \right)^{-1}  \ .
\end{equation}
Herein~$\theta$ is the fluids temperature, $\alpha^{th}_{fl}$ is the thermal
volumetric expansion coefficient and~$\rho_{fl_{ref}}$ is the mass density at a reference temperature~$\theta_{ref}$. Parameters are given in~\cite{Silicon_Oil} 
 and
Table~\ref{tab:MatPar_fluid_density}.
\begin{table}[ht]
 \centering
  \caption{Material parameters for Eq.~\eqref{eq:fluid_density}} 
  {\begin{tabular}{p{1.6cm}p{2.0cm}p{1.6cm}p{1.4cm}}
    \hline & & & \\[-3mm]
    parameter & value & parameter & value\\
    \hline & & & \\[-2mm]
        $\ \ $ $\rho_{fl_{ref}}$               & $0.963\,$ g $\cdot$ cm$^{-3}$
    &   $\ \ $ $\theta_{ref}$           & $298.13 \ \rm $K   \\[2mm]
        $\ \ $ $\alpha^{th}_{fl}$       & $9.4\cdot10^{-4} \ \rm K^\inv$ 
    &   \\[1mm]
    \hline
  \end{tabular}}
  \label{tab:MatPar_fluid_density}
\end{table}  

\begin{figure}[ht]
	\centering
  \includegraphics[width=0.5\textwidth]{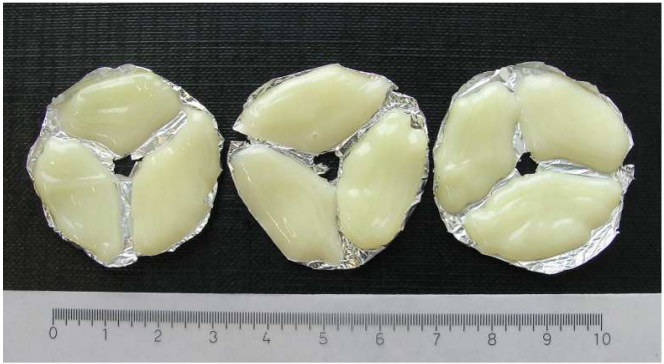}
	\caption{Three chosen specimen for the measurement of volume shrinkage during the curing reaction (dimension:~$[$mm$]$).}
	\label{fig:shrinkage_specimen}
\end{figure}

Since quasi-isothermal experiments should be run and the observed process is an
exothermic reaction, not only the experimental setup but also the sample itself
need to be specifically adapted. Toward this end the resin specimen was split into three partial volumes which feature an appropriate ratio of surface to volume. Consequently, the generated heat will be conducted to the silicon oil
more efficiently. Fig.~\ref{fig:shrinkage_specimen} shows three exemplary chosen specimen.

In addition, the sample holder was coated with the silicone
oil prior to each measurement, which should allow a free
volume shrinkage without the build-up of residual stresses. The amount of the
total sample volume was chosen in a way that almost the entire signal range of
the capacitive sensor is utilized during the curing reaction. The average sample weight out of 25 conducted experiments (five repititions per temperature level) was $m=1.61 \pm 0.15\,$g. Upon immersion of
the prepared sample into the testing liquid, a rotational movement from the vertical to the
horizontal is performed since the formation of interfering air bubbles should be
avoided. No evidence of air bubbles was found during the measurements, in the fully cured samples or within the results.


\subsection{Measurement characteristics and preliminary experiments}
\label{sec:Preliminary_experiments}


Based on the measuring task some considerations on the concomitant requirements to the experimental setup are given hereunder. Comparable to the DSC (cf. section~\ref{sec:DSC}) the specimen needs to be as small as possible in order to diminish the impact of the reaction heat and to achieve an homogeneous temperature distribution in the specimen. 
%
%
%
Consequently, both a small specimen mass as well as a sample shape allowing for good heat transport (see Fig.~\ref{fig:shrinkage_specimen}) minimises the heating related error of the experimental method.

Naturally, there are limitations to reducing the mass of the specimen due to measurement precision of the experimental setup.
During the development process of the test rig a specimen mass of $2\,$g was found to be adequate in order to fulfil the measuring task. Since the density of specimen and fluid for immersion are roughly the same with a value of approximately~$1\,$g$\cdot$cm$^{-3}$, and the shrinkage of epoxy resin is in a range of about~$5\,\%$, it can be estimated that a change in force of about~$1\,$mN has to be measured. In order to resolve this change into at least $100$~steps, the long-term stable measurement resolution should be at least~$10\,\mu$N. This theoretical considerations can be compared to results obtained from specific preliminary experiments which analyse the characteristics of the presented experimental setup. 

For this reason, the conversion factors for the sensors were determined first. These factors convert a signal change of either the capacitive sensor or the strain gauges into an equivalent force. There was no detectable difference between factors for the measuring pointer on the one hand and the reference pointer on the other hand. Hence, there is one factor for each sensor type. The factors can be determined from the equation
\begin{equation}
\label{eq:signal_conversion}
  \ \  \Delta F = K_i \cdot \Delta S_i \, , \ \ i=sg, cap ,
\end{equation} 
where $\Delta F$, $K_{i}$ and $\Delta S_i$ are equivalent force change, conversion factor and corresponding signal change, respectively. The identified values are given in Table~\ref{tab:conversion_factors}.
\begin{table}[ht]
 \centering
  \caption{Conversion factors for converting the signal changes of strain gauges~(sg) and capacitive sensors~(cap) into an equivalent force} 
  {\begin{tabular}{p{0.7cm}p{2.5cm}p{0.7cm}p{1.8cm}}
    \hline & & & \\[-3mm]
    factor & value & factor & value\\
    \hline & & & \\[-2mm]
        $\, $ $K_{sg}$               & $0.553\,$mN$\cdot$V$\cdot\mu$V$^{-1}$
    &         $K_{cap}$              & $0.164 \ \rm$mN$\cdot$V$^{-1}$   \\[1mm]
    \hline
  \end{tabular}}
  \label{tab:conversion_factors}
\end{table}  

\begin{figure}[ht]
	\centering
  \includegraphics[width=0.5\textwidth]{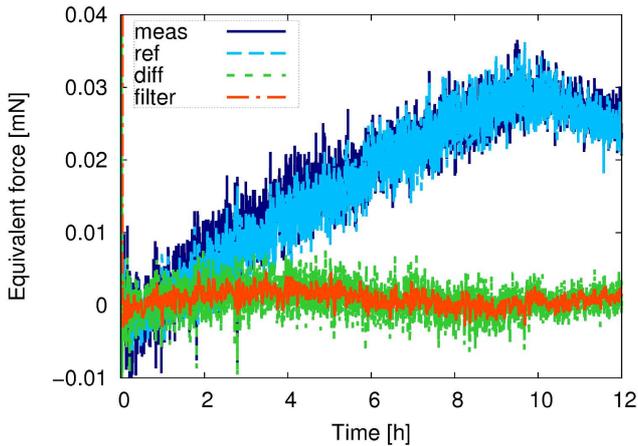}
	\caption{Long-term stability and resolution of the capacitive sensors: by employing both the compensation principle (difference between measuring and reference signal) as well as the mathematical filtering of oscillations (smoothing) the theoretically constant signal stays within a measuring corridor of~$10\,\mu$N (converted from voltage). This value coincides the required resolution based on theoretical considerations (see text).}
	\label{fig:system_free}
\end{figure}

Next, an assessment on resolution and stability can be achieved by a load free measurement over a comparably long time interval only subjected to environmental influences as depicted in Fig.~\ref{fig:system_free}. Here, only the capacitive sensors are considered, since they are responsible to measure the long-term stable signal (cf. section~\ref{sec:Principle_measurement_and_exp_setup}).
It shows four graphs measured over a period of twelve hours with the sampler being immersed but not containing a specimen. One signal is obtained from the measuring pointer (meas) and one from the reference pointer (ref) --- (compare the experimental setup in section~\ref{sec:Principle_measurement_and_exp_setup}). These two are subjected to environmental influences, which cause oscillations and long-term deviations from the theoretically constant value (drift) of about~$41\,\mu$N. Building the difference between these two signals the environmental effects nearly vanish from the signals. The corresponding graph~(diff) varies in a range of approximately~$16\,\mu$N. Additionally, a mathematical filter with Gaussian distribution of the weighting factors within a time range of~$\pm \, 45\,$s around the supporting point is applied. The resulting slightly smoothed signal~(filter) now exhibits variations of~$10\,\mu$N which corresponds the desired resolution based on the theoretical assumptions presented in the text above. This behaviour could be demonstrated in multiple experiments with partially even longer time periods for measurement. Hence, the experimental setup is suitable to measure the volume shrinkage since the demanded resolution and long-term stability are given.

\begin{figure}[ht]
	\centering
  \includegraphics[width=0.5\textwidth]{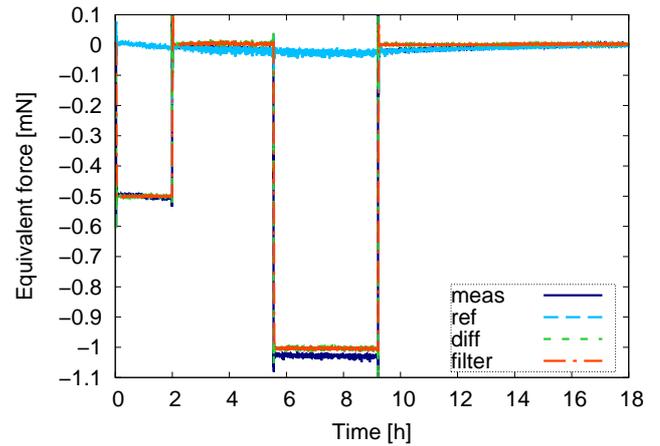}
	\caption{Long-term behaviour of the test rig with temporary applied constant loads: the setup shows linear behaviour, the compensation of environmental influences (like temperature changes and air flow) works properly and the mathematical filter (smoothing) reduces the overshooting effects.}
	\label{fig:system_loads}
\end{figure}

Fig.~\ref{fig:system_loads} shows the long-term behaviour of the test rig without the sampler beeing attached to it. Two different constant forces have been applied temporary by placing small weights~($50\,$mg, $100\,$mg) on the measuring pointer. Linear behaviour is given and the compensation of environmental influence works properly and as intended. The mathematical filter reduces the overshooting at the time points with abrupt signal changes.

It can be concluded, that the measuring properties were proved to show a long-term stable behaviour for at least twelve hours (cf.~Fig.~\ref{fig:system_free} and \ref{fig:system_loads}), which is sufficient since the conducted experiments only lasted four hours. 

Further preliminary experiments have been con\-duct\-ed. For one of them the measurement system was completely blocked. It proved the signal drift of the capacitive sensors to be one magnitude smaller than the desired resolution. Hence, this effect is negligible. Swelling tests had been conducted at measurement conditions and did not show any influence to the specimen. The viscosity of the damping fluid had been adjusted in a way that there is no viscous behaviour superimposing the volume shrinkage of the adhesive (this time-dependent effect depends on the scale of the investigated force differences). Actually, the viscosity of the employed silicon oil was decreased to such an extent that there is now a fading oscillation on the onset of an abrupt load change over a short period of time (compared to the curing time) which, however, is taken into account by the mathematical filter (smoothing). This is one aspect within the following section~\ref{sec:Processing_Shrinkage}.

%


\subsection{Procedural protocol and calculations}
\label{sec:Processing_Shrinkage}



It is essential to work very accurate and follow a specific procedural protocol to calculate the density or volume of the specimen from the data sets. The proposed procedure reads as
follows:
\begin{enumerate}
 \item Set up reference state: $S_{sg}=0$, $\STAPEL S!^\SLtilde_{sg}=0$,
$\STAPEL S!^\SLtilde_{cap}=0$
 \item Start measurement with time zero: $t=t_0$
 \item Begin adhesive application i.e. mixing the two components of the resin:
$t=t_1$
 \item Measure weight of specimen: $m_s$
 \item Place specimen on sampler and attach it to the experimental setup: $t=t_2$
 \item Positioning of the capacitive sensor at the measuring beam: $t=t_3$,
$S_{cap}(t_3)=0$
 \item ... curing reaction for the desired time range ($4h$) ...
 \item Remove sample and measure resulting signal change at measuring beam:
$t=t_e$, $S_{abs}(t_e)= \qquad\qquad\qquad$ $S_{sg}(t=t_e+0) - S_{sg}(t=t_e-0)$
\end{enumerate}
Herein,~$S$ and $\STAPEL S!^\SLtilde$ are the measuring signal from the measuring and the reference pointer, respectively. Abbreviations are used for indexing of strain gauges~($sg$), capacitive sensors~($cap$) and absolute value~($abs$).

It is essential to point out that at the beginning of each experiment the specimen has ambient temperature~($25\,^\circ$C) independant of the experimental temperature. This is due to practical reasons. Since the complete setup is not subjected to this experimental temperature there will always be a period of time while prepairing the specimen at whichs end its current temperature and degree of cure are undefined. This is specifically problematic for higher temperature levels where the curing reaction proceeds very fast. Accordingly the initial state of the specimen by terms of temperature and degree of conversion is very well defined by the applied method. This is also taken into account in the modelling (see section~\ref{sec:model_volume_changes}).
%
%
 Nevertheless, it is yet not that well defined as the state of the curing reaction and specimen temperature at the end of the experiment (after four hours), since the latter has homogenised due to the former being subsided. Hence, this point in time is used to determine the absolute value of the specimen density as can be seen from the last bullet of the procedural protocol.

 As indicated in
section~\ref{sec:Principle_measurement_and_exp_setup} there are two types of
sensors with different properties at each pointer. In order to finally obtain the desired results for both the absolute density values as well as its affiliated, comparably small changes due to thermochemical processes the specific advantages of each measuring system has to be extracted. 

On the one hand the strain gauges can monitor absolute values but is subjected to long-term signal drift. Its stability is not given within the magnitude desired to detect.
On the other hand it is shown in section~\ref{sec:Preliminary_experiments} that the capacitive sensors comprise the required resolution and long-term stability.
However, an optimal resolution of the curing graph can only be achieved if their measuring range of about~$0.3\,$mm is fully exploited by the pointers movement only due to curing related volume changes.
That is of course at the same time, the measuring range is exceeded by movement of the pointer due to the buoyancy force upon immersion of the specimen itself (since volume changes only cover~$5\,\%$ of the absolute displaced volume).
To overcome this insufficiency the principle of the work of~\cite{Nebel2009} 
 is applied.
It is based on the idea that the measuring pointer upon the sample being attached immerses into the measuring range of the capacitive sensor such that it can now be employed to monitor the smaller volume changes due to curing.
Virtually, this means the capacitive sensors position is adjusted to the appropriate distance after the sample is attached (compare to bullet six of the procedural protocol). 
After four hours the final value of its signal is set to be the reference value of the course (set to zero). 
It now functions as supporting point for the addition of the corresponding equivalent forces of the absolute value from the strain gauges signal (also measured after four hours) and the associated relative signal change from the capacitive signal.

In order to better 
understand the principle of evaluation of the measuring signals a short
theoretical deduction for the calculation of the specimens density is given
hereunder. Starting point is the comparison of the measuring signal (regardless
of which measurement system is considered) with the specimen being on the
sampler or not. This signal difference is equivalent to the force difference between
the samples weight and buoyancy (for conversion factors see
section~\ref{sec:Preliminary_experiments}).
\begin{equation}
\label{eq:equilibrium_forces}
  \ \ F_{res}(t) = F_{meas}(t) = F_g - F_b(t) \ .
\end{equation}
Using a precision balance and assuming constant mass the gravity force~$F_g$ is
easily determined with a constant gravitation of~$g=\,9.81\,$m$\cdot$s$^{-2}$. In
section~\ref{sec:Preliminary_experiments} it was indicated that no changes in sample
mass due to swelling could be detected. The measured force is a sum of an
absolute value taken from the strain gauge at a certain point in time~$t_x$ and
a relative value obtained from the capacitive sensors.
\begin{equation}
\label{eq:measured_force}
  \ \ F_{meas}(t) = F_{abs}(t_x) + F_{rel}(t)  \ .
\end{equation}
As mentioned above the
absolute density can not reliably be determined at the beginning of the
experiment due to practical handling. Hence, the value of the strain gauge at
the end of the experiment~$S_{abs}(t_e)$ after four hours ($t=t_e=t_x$) is used
to calculate the absolute density of the specimen. That means the sample needs
to be removed from the sampler due to signal changes owing to curing and
drifting (compare bullet eight of the procedural protocol). Taking into account that the signal difference of the capacitive
sensors~$S_{cap}(t)$ needs to be shifted by the value reached after four
hours~$S_{cap}(t_e)$ 
it yields
\begin{equation}
\label{eq:measured_signals}
  \ \ F_{meas}(t) = S_{abs}(t_e) \cdot K_{sg} + (S_{cap}(t_e) - S_{cap}(t)) \cdot
K_{cap} \ ,
\end{equation}
where~$K_{sg}$ and~$K_{cap}$ are the factors presented in
section~\ref{sec:Preliminary_experiments}. Assuming the sample volume~$V_s$ to
be identical to the volume of the displaced fluid at each time, the volume of
the specimen can be calculated via
\begin{equation}
\label{eq:sample_volume}
  \ \ V_s(t) = \dfrac{1}{\rho_{fl}(\theta)} \left( m_s - \dfrac{1}{g} \cdot
F_{meas}(t) \right)  \ .
\end{equation}
Herein~$\rho_{fl}(\theta)$ is the temperature dependent fluid density (given by Eq.~\eqref{eq:fluid_density}) and~$m_s$
the mass of the specimen. It follows for the sample density
\begin{equation}
\label{eq:sample_density}
\begin{array}{lcl}
  \ \ \rho_s(t)
    = \rho_{fl}(\theta) \left( 1 - \dfrac{1}{m_s g} \cdot F_{meas}(t)
\right)^{-1}  \\[5mm]
  \ \ \phantom{\rho_s(t)}
    =  \rho_{fl}(\theta) \left( 1 - \dfrac{F_{meas}(t)}{F_g}  \right)^{-1} \ .
\end{array}
\end{equation}
Evaluation of experimental data following Eq.~\eqref{eq:sample_density} is achieved by macros. Off-set values, smoothing and averaging is considered there. The results are compared to simulations and presented including the modelling approach within the next section~\ref{sec:model_volume_changes}.


\subsection{Results of the experiments and simulation}
\label{sec:model_volume_changes}

\begin{figure}[ht]
	\centering
  \includegraphics[width=0.5\textwidth]{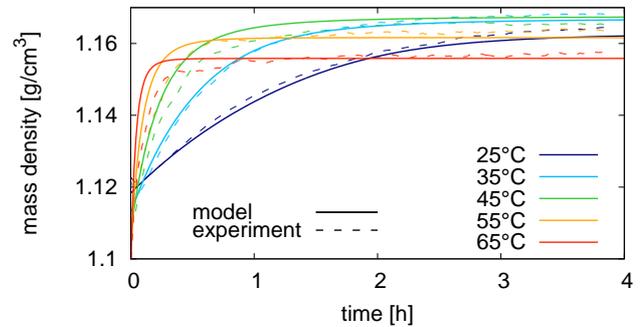}
	\caption{Volumetric changes due to curing: comparison of experimental data (dashed) and simulation data for different time-temperature regimes. Both graphs are obtained after numerical treatment of Eq.~\eqref{eq:sample_density} and~\eqref{eq:model_density}.}
	\label{fig:results_shrinkage_time}
\end{figure}

The experimental setup which is described and characterised within the previous sections~\ref{sec:Principle_measurement_and_exp_setup} to~\ref{sec:Processing_Shrinkage} is employed to measure the adhesives chemically related volume shrinkage at certain isothermal temperature levels, which are~$25\,^\circ$C, $35\,^\circ$C, $45\,^\circ$C, $55\,^\circ$C and $65\,^\circ$C. The curing reaction of the investigated material is recorded for four hours. This was found to be a time range after which the curing is completed or at least has turned into a diffusion controlled reaction (cf. Fig~\ref{fig:curing}). Five measurements at each temperature are conducted after the procedural protocol given in section~\ref{sec:Processing_Shrinkage}. Subsequently, the data is subjected to the mathematical filtering (with Gaussian distribution) as described in section~\ref{sec:Preliminary_experiments}. Finally, the average of the five data sets per temperature level is formed and depicted in Fig.~\ref{fig:results_shrinkage_time} (dashed lines).


%
Based on the modelling approach for the progress of the chemical reaction deduced in section~\ref{sec:DoC} an ansatz describing the temperature and degree of cure dependent volumetric changes is formulated. It reads as follows:
\begin{equation}
\label{eq:model_relative_volume}
  \ \ J_{\theta C} (\theta, q) = \dfrac{\widetilde{\varrho}}{\varrho} = {\rm exp}\left[ \alpha_\theta \, \big(\,\theta - \tilde{\theta}\,\big) + \beta_q \, q\right] \ .
\end{equation}
Dissolved for the specimens density it yields:
\begin{equation}
\label{eq:model_density}
  \ \ {\varrho} (\theta, q) = \widetilde{\varrho} \cdot {\rm exp}\left[ \alpha_\theta \, \big(\,\tilde{\theta} - \theta \,\big) - \beta_q \, q\right] \ .
\end{equation}
Herein, $\widetilde{\varrho}$ and $\varrho$ is the reference and current specimen density, respectively. The model parameters are~$\alpha_\theta$ and $\beta_q$, which can be interpreted as a volumetric heat expansion coefficient and the maximum volumetric chemical shrinkage.
%
%
Within Eq.~\eqref{eq:model_relative_volume} and~\eqref{eq:model_density} the functions~$\theta$ and~$q$ are time dependent. The former can be included from any arbitrary process and the latter (also dependent on this temperature) is numerically calculated on the basis of Eq.~\eqref{eq:q_ansatz}.
%
%
Numerical treatment of the analytical models and the data processing for Eq.~\eqref{eq:q_from_exp} and~\eqref{eq:sample_volume} or~\eqref{eq:sample_density} is not considered within this article.
Parameter identification was conducted via the solution of a least squares problem through nonlinear optimization techniques. The corresponding results are given in Table~\ref{tab:MatPar_Volume}.
%
%
\begin{table}[ht]
 \centering
  \caption{Material parameters for Eq.~\eqref{eq:model_density}} 
  {\begin{tabular}{p{1.6cm}p{2.0cm}p{1.6cm}p{1.4cm}}
    \hline & & & \\[-3mm]
    parameter & value & parameter & value\\
    \hline & & & \\[-3mm]
        $\ \ $ $\thetatil$           & $295 \ \rm K$ 
    &                                   &   \\
        $\ \ $ $\alpha_\theta$  & $5\cdot10^{-4} \ \rm K^\inv$ 
    &  $\ \ $ $\beta_q$          & $-0.053$    \\[1mm]
    \hline
  \end{tabular}}
  \label{tab:MatPar_Volume}
\end{table}

Fig.~\ref{fig:results_shrinkage_time} depicts the results of the real-time development of the specimens density due to the curing reaction. The experimental curves (dashed) and the graphs from the simulation follow upon numerical treatment of the Eq.~\eqref{eq:sample_density} and~\eqref{eq:model_density}, respectively.

It can be seen from Fig.~\ref{fig:results_shrinkage_time} that the density at the end of the experiment does not line up in a linear manner. This could be expected due to the different temperature levels. The solution to this can be found in Fig.~\ref{fig:curing} on page~\pageref{fig:curing}. Since, the curing turns into a diffusion controlled reaction when the materials glas transition temperature increases the current experimental temperature, the degree of cure stagnates at a certain value. For this reason, the specimens shrinkage will stop at an intermediate state. Hence, the density for the samples cured below~$45\,^\circ$C is not fully developed in the sense of a completed chemical reaction.

%
%
%

%
\section{Summary}
\label{sec:Conclusions}

%
%
%

%

Within this article, the experimental determination and modelling of thermochemical volume changes of an epoxy based adhesive is presented. The main focus is the especially constructed experimental setup based on Ar\-chi\-me\-des principle. The parallel construction for the compensation of environmental influences and the combination of two different sensor systems in order to determine both the absolute sample density as well as small changes due to chemical shrinkage with an appropriate resolution are two characteristic features and proved to function very well. A procedural protocol and corresponding calculations are provided to enable the reader to comprehend measurement, subsequent data processing and evaluation of the density. Due to the comparative simulation of the volume shrinkage process, the curing reaction itself has to be captured. differential scanning calorimetry (DSC) are employed to define the progress of the exothermic reaction. Phenomenological models to represent reaction kinetics and the thermochemically related isotropic volume changes are given. For purposes of comparability with the experiments, material parameters are identified and the results are depicted. 

In the future, different material systems will be studied with the presented experimental setup. Toward this end, the extension to higher temperature ranges represents an important aspect.


\begin{acknowledgements}
This  research  is  supported  by  the  German Research Foundation 
 (DFG) within the Collaborative  Research  Centre/Transregio  39 
PT-PIESA. This support is greatly acknowledged.
\end{acknowledgements}

\bibliographystyle{spmpsci}      
\bibliography{lit_Archimedes_Principle_Rudolph_Etal}   

\begin{thebibliography}{10}
\providecommand{\url}[1]{{#1}}
\providecommand{\urlprefix}{URL }
\expandafter\ifx\csname urlstyle\endcsname\relax
  \providecommand{\doi}[1]{DOI~\discretionary{}{}{}#1}\else
  \providecommand{\doi}{DOI~\discretionary{}{}{}\begingroup
  \urlstyle{rm}\Url}\fi

\bibitem{DP410_2003}
{3M Deutschland}: 3M$^{\rm TM}$ Scotch-Weld$^{\rm TM}$ EPX Epoxy Adhesive DP410
  - Technical datasheet.
\newblock {3M Deutschland GmbH} (2003)

\bibitem{DiBenedetto_1987}
DiBenedetto, A.T.: {Prediction of the glass transition temperature of polymers:
  A model based on the principle of corresponding states}.
\newblock J Polym Sci Pol Phys \textbf{25}(9), 1949--1969 (1987)

\bibitem{Drossel_Etal_2009_CIRP}
Drossel, W.G., Hensel, S., Kranz, B., Nestler, M., Goeschel, A.: Sheet metal
  forming of piezoceramic-metal-laminar structures -- simulation and
  experimental analysis.
\newblock Cirp Ann-Manuf Techn \textbf{58}(1), 279 -- 282 (2009)

\bibitem{Fournier_Etal_1996}
Fournier, J., Williams, G., Duch, C., Aldridge, G.A.: {Changes in Molecular
  Dynamics during Bulk Polymerization of an Epoxide-Amine System As Studied by
  Dielectric Relaxation Spectroscopy}.
\newblock Macromolecules \textbf{29}(22), 7097--7107 (1996)

\bibitem{Silicon_Oil}
GmbH, W.C.: Wacker Silicone Fluids AK - Technical datasheet.
\newblock Wacker Chemie AG (2002)

\bibitem{Habenicht2006}
Habenicht, G.: {Kleben}.
\newblock Springer Verlag (2006)

\bibitem{Halley_Mackay_1996}
Halley, P.J., Mackay, M.E.: {Chemorheology of thermosets - an overview}.
\newblock Polym Eng Sci \textbf{36}(5), 593--609 (1996)

\bibitem{Holst2001}
Holst, M.: {Reaktionsschwindung von Epoxidharz-Systemen}.
\newblock Dissertation, Technische Universit\"{a}t Darmstadt (2001)

\bibitem{Hulder2008}
H\"{u}lder, G.: {Zur Aush\"{a}rtung kalth\"{a}rtender Reaktionsharzsysteme
  f\"{u}r tragende Anwendungen im Bauwesen}.
\newblock Ph.D. thesis (2008)

\bibitem{Khoun2010}
Khoun, L., Hubert, P.: {Cure Shrinkage Characterization of an Epoxy Resin
  System by Two in Situ Measurement Methods}.
\newblock Polymer Composites \textbf{31}(9), 1603--1610 (2010)

\bibitem{Kolmeder_Lion_2010}
Kolmeder, S., Lion, A.: {On the thermomechanical-chemically coupled behavior of
  acrylic bone cements: Experimental characterization of material behavior and
  modeling approach}.
\newblock Technische Mechanik \textbf{30}(1-3), 195--202 (2010)

\bibitem{Landgraf2014}
Landgraf, R., Rudolph, M., Scherzer, R., Ihlemann, J.: {Modelling and
  simulation of adhesive curing processes in bonded piezo metal composites}.
\newblock Computational Mechanics  (2014)

\bibitem{Li2004}
Li, C., Potter, K., Wisnom, M.R., Stringer, G.: {In-situ measurement of
  chemical shrinkage of MY750 epoxy resin by a novel gravimetric method}.
\newblock Composites Science and Technology \textbf{64}(1), 55--64 (2004)

\bibitem{Lion_Yagimli_2008}
Lion, A., Yagimli, B.: {Differential scanning calorimetry - continuum
  mechanical considerations with focus to the polymerisation of adhesives}.
\newblock Z Angew Math Mech \textbf{88}(5), 388--402 (2008)

\bibitem{Mahnken_2013}
Mahnken, R.: {Thermodynamic consistent modeling of polymer curing coupled to
  visco–elasticity at large strains}.
\newblock Int J Solids Struct \textbf{50}(13), 2003--2021 (2013)

\bibitem{Nawab2013}
Nawab, Y., Shahid, S., Boyard, N., Jacquemin, F.: {Chemical shrinkage
  characterization techniques for thermoset resins and associated composites}.
\newblock Journal of Materials Science \textbf{48}(16), 5387--5409 (2013)

\bibitem{Nebel2009}
Nebel, S.: {Ein Beitrag zur experimentellen und numerischen Analyse
  zeitabh\"{a}ngiger Eigenschaften von DMS-Messstellen}.
\newblock Ph.D. thesis, Technische Universit\"{a}t Chemnitz (2009)

\bibitem{DaSilva2011}
da~Silva, L.F.M., \"{O}chsner, A., Adams, R.D.: {Handbook of Adhesion
  Technology}.
\newblock Springer Verlag (2011)

\bibitem{Wenzel2005}
Wenzel, M.: {Spannungsbildung und Relaxationsverhalten bei der Aush\"{a}rtung
  von Epoxidharzen}.
\newblock Ph.D. thesis, TU Darmstadt (2005)

\end{thebibliography}



\end{document}